# Experimental investigation of quantum key distribution over water channel

Shicheng Zhao, Wendong Li* , Yuan Shen, YongHe Yu, XinHong Han, Hao Zeng, Maoqi Cai, Tian Qian, Shuo Wang, Zhaoming Wang, Ya Xiao and Yongjian Gu*

*Department of Physics, Ocean University of China, Qingdao 266100, China.*

E-mail: : yjgu@ouc.edu.cn, liwd@ouc.edu.cn

***Abstract:*** *Quantum key distribution (QKD) has undergone significant development in recent decades, particularly with respect to free-space (air) and optical fiber channels. Here, we report the first proof-of-principle experiment for the BB84 protocol QKD over a water channel. Firstly, we demonstrate again the polarization preservation properties of the water channel in optical transmission according to the measured Mueller matrix, which is close to the unit matrix. The reason for the polarization preservation, revealed by Monte Carlo simulation, is that almost all the received photons are unscattered. Then, we performed the first polarization encoding BB84 protocol QKD over a 2.37m water channel. The results show that QKD can be performed with a low quantum bit error rate (QBER), less than 3.5%, with different attenuation coefficients.*

***OCIS* codes:** (060.5565)Quantum communications; (270.5568) Quantum cryptography; (010.4450)Oceanic optics; (010.0010)Atmospheric and oceanic optics.

## 1. Introduction

Quantum key distribution (QKD)[1–3] can provide theoretical communication security for distant parties. Since the BB84 protocol, the first practical QKD protocol, was proposed by Bennett and Brassard[4] in 1984, QKD technology has been developed both theoretically and experimentally. Several protocols[5–8], such as E91 [5] and B92 [6], have been proposed. Decoy states[9– 11] and the Measurement-Device-Independent (MDI) protocol [12] were put forward to solve security problems caused by the imperfection of single-photon sources and detectors, respectively. QKD has been experimentally demonstrated over both free-space (air)[13] and optical fiber channels[14, 15]. The first free-space QKD experiment [16] was carried out in 1989 over a distance of 32 cm; more recently, satellite-based polarization encoding QKD with a decoy state has been achieved from satellites to the ground over distances of up to 1200 km[17]. The QKD was also achieved over 404 km of standard telecom fiber[18], representing a great increase from 1 km in 1993[19].

Besides the air and optical fiber channels, seawater is also vital for communications. Underwater communication plays an important role in tactical surveillance, pollution monitoring, undersea explorations, climate monitoring, and oceanography research. Underwater QKD, which can ensure the security of underwater communications, was proposed by Lanzagorta[20, 21] in 2010 through a theoretical analysis of its feasibility. The performance of underwater QKD was further studied by Shi et al. [22] using vector Monte Carlo simulations. Ji et al.[23] experimentally demonstrated that polarization states and polarization entangled states can survive well in a 3m Jerlov type I seawater channel. Bouchard et al.[24] experimentally studied the properties of a twisted photon over a 3 m water channel with turbulence and calculated the quantum bit error rate (QBER) and bit rate according to probability-of-detection matrices measured during the experiment. These researches have demonstrated the feasibility of underwater quantum communication, especially the QKD. Naturally, the underwater QKD should be done experimentally and developed for the practical application. However, up to now, there is no report on the experimental QKD over a water channel. The BB84 protocol, as the basic QKD protocol, is chosen to be realized underwater here

In this study, we performed a proof-of-principle experiment for BB84 protocol underwater QKD with polarization encoding over a 2.37m simulated seawater channel. Firstly, we investigated the polarization preservation properties of the channel by measuring the Mueller matrix with single photons. The measured Mueller matrix is close to unit matrix, and this indicates that the channel can preserve the polarization well. Using the Monte Carlo method, we find that almost all the received photons are unscattered, and so the polarization of received photons are nearly unchanged. Then, we performed the polarization encoding BB84 protocol QKD over the channel. The results show that QKD over the water channel can be performed with a low QBER (less than 3.5%) for different attenuation coefficients. The results demonstrate that polarization encoding BB84 protocol QKD is feasible over water channel.

## 2. Polarization preservation properties of the water channel

Underwater QKD can provide theoretical security for underwater communications. We considered a realization of underwater QKD using the polarization of the photon[22, 23], whose performance depends on the polarization preservation properties of the water channels. Seawater is complex and its optical properties will also be different. As the polarization preservation properties of the channel is the premise for polarization encoding QKD, we studied the polarization preservation properties of the seawater at first. Similar to the description of an optical element, the evolution of the polarization state in water channels can be expressed as a 4×4 Mueller matrix [25]. To examine how the polarization changes in water channels, we performed an experiment to measure the Mueller matrix of the water channel with single photons. The experimental setup, as shown in Fig. 1, included the transmitter, seawater channel, and receiver. Considering the absorption and scattering of photons by sea water, the laser we used is 488nm which is within the blue-green optical window of seawater (430 to 570nm[26]). The polarizer (P1 ) and quarter-wave plate (Q1 ) at the transmitter part were used to generate different input states. These generated states were further passed through a 2.37m sea water channel whose Mueller matrix needs to be determined. The polarization properties of the corresponding output states were analyzed by using a quarter-wave plate (Q2) and an analyzer (P2) whose transmission axis is kept parallel to P1 . The water was prepared by adding inorganic salt to ultra-pure water with a resistivity of 18MΩ/cm. According to the major constituents of seawater[27], the salt was made up of $NaCl$, $MgCl_2$, $Na_2SO_4$ , and $CaCl_2$ in concentrations of 24.53, 5.20, 4.09, and 1.16 g/L, respectively. After adding about 1g $Al(OH)_3$ to the 191L water, the absorption coefficient was 0.117/m, and the attenuation coefficient was 0.683/m. The absorption coefficient and attenuation coefficient (the sum of the absorption coefficient and scattering coefficient) were measured by AC-S Sea-Bird Scientific.

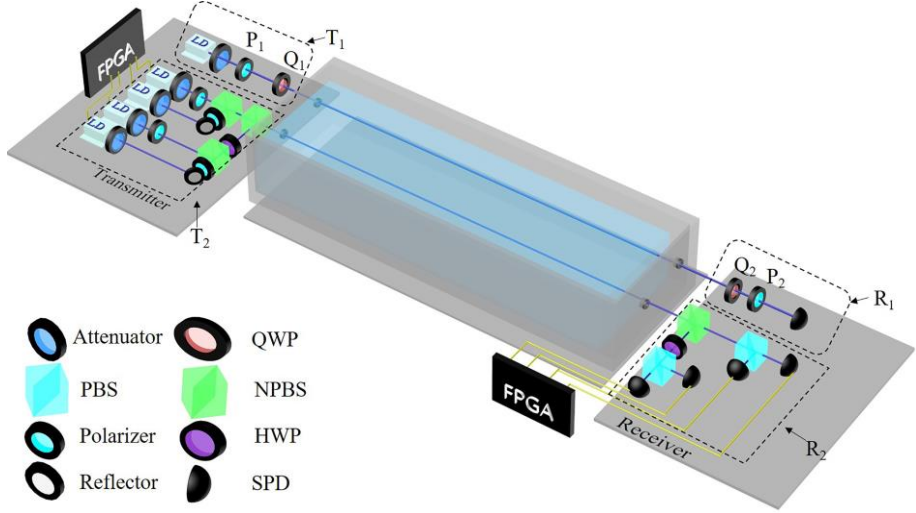

**Fig. 1.** (color online) Sketch of the Mueller matrix measurement (T1 and R1) and underwater BB84 protocol QKD (T2 and R2). HWP represents the half-waveplate, QWP represents the quarter-waveplate NPBS represents the non-polarizing beam splitter, and PBS represents the polarizing beam splitter.

The photon's polarization state can be represented as a Stokes vector S, and the evolution of the state is described as:

$$S_1 = M \cdot S_0, \qquad (1)$$

where S0 and S1 are the Stokes Vectors of the incident and output light, respectively[25, 28, 29], and $M = M_{p_2} \cdot M_q(\theta_2) \cdot M_w \cdot M_q(\theta_1) \cdot M_{p_1}$ is the Mueller matrix for the system. $M_w$ is the Mueller matrix of the water channel to be measured, $M_{p_1}$ and $M_{p_2}$ are the Mueller matrices of the polarizer (horizontal for our experiment), and $M_q(\theta_1)$, $M_q(\theta_2)$ are the Mueller matrices of the quarter waveplates at the transmitter and receiver, respectively, and have the form[31]:

$$M_{p_1} = M_{p_2} = \frac{1}{2} \begin{pmatrix} 1 & 1 & 0 & 0 \\ 1 & 1 & 0 & 0 \\ 0 & 0 & 0 & 0 \\ 0 & 0 & 0 & 0 \end{pmatrix}, \qquad (2)$$

$$M_q(\theta_i) = \frac{1}{2} \begin{pmatrix} 1 & 0 & 0 & 0 \\ 0 & a & b & d \\ 0 & b & c & -e \\ 0 & -d & e & cos(\delta) \end{pmatrix}, \quad (3)$$

where $a = cos^2(2\theta_i) + cos(\delta) sin^2(2\theta_i)$ $b = cos(2\theta_i) sin(2\theta_i) - cos(2\theta_i) cos(\delta) sin(2\theta_i)$, $c = cos(\delta) cos^2(2\theta_i) + sin^2(2\theta_i)$ $d = sin(2\theta_i) sin(\delta)$, $e = cos(2\theta_i) sin(\delta)$; i =1 or 2, and $\theta_1$, $\theta_2$ represent the angle of the fast axis of the waveplate at the transmitter and receiver, respectively. The $\delta$ is the phase difference between the fast and slow axis, which is $\pi/2$ for a QWP. Since there are 16 unknown elements in $M_w$, one needs to choose four different input states to obtain the Mueller matrix. Here, we generated four polarization states by rotating $\theta_1$ to 0, $\pi/8$, $\pi/4$, and $3\pi/8$ at the transmitter, and the average photon number was attenuated to about 0.2 per pulse when $\theta_1$ was 0. At the receiver, $\theta_2$ was set to 0, $\pi/8$, $\pi/4$, and $3\pi/8$ for each polarization state and the photon number was recorded. Thus, we obtained 16 sets of data (photon number) which indicate the first item of the Stokes

Vector. Then, a system of linear equations of 16 unknowns elements of $M_w$ was set up according Eq. 1, and $M_w$ was obtained by solving it. In this study, we mainly investigated the polarization preservation properties of the water channel and the measured Mueller matrix was normalized. As shown in Fig. 2, the resulting Mueller matrix is almost a unit matrix, where the minimum value of diagonal elements is larger than 0.976, and the maximum absolute value of non-diagonal elements is smaller than 0.093. This means that the water channel can preserve all of the polarization states well. In order to further verify our results, we also calculated the fidelity[30] of the polarization state after passing through the water channel. The mean fidelity of four polarization states (horizontal, vertical, $45°$ linear polarization, and $135°$ linear polarization) is as high as 0.9823±0.0048.

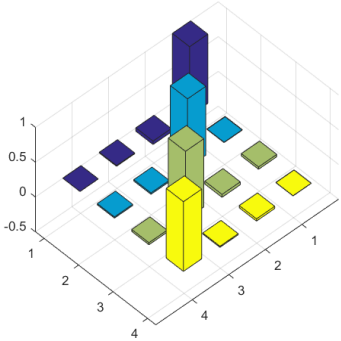

**Fig. 2.** (color online ) Mueller matrix of the water channel measured by single photons with a distance of 2.37m

To explain why the water channel can preserve the polarization, we calculated the ratio of the number of scattered photons to the number of |Rthe total received photons using the Monte Carlo method [22]. During the simulation, we made three improvements based on the open source MCML program[32]: (1) simulating the Gaussian beam instead of the ideal beam with infinity narrow size according to the mathematical model in Ref. [33]; (2) introducing a two term Henyey-Greenstein function[34] to obtain the scattering angle; (3) considering the optical properties at the receiver, including aperture and field of view

(FOV). For the simulated seawater channel (the absorption coefficient was 0.117/m, the attenuation coefficient was 0.683/m), we find that only about 0.16% of the received photons are scattered, others are unscattered at a distance of 2.37m when the aperture, FOV are 2.54cm and 10◦ , respectively. The polarization states of the unscattered photons remained unchanged. Given the limitations of aperture and FOV, most of the received scattered photons had small scattering angles. Scattered photons with small scattering angles will have little effect on polarization states, as shown theoretically by [23] and experimentally by [25]. This is the main reason why the Mueller matrices of seawater are close to the unit matrix, and why polarization can be preserved for QKD over seawater channels.

We also performed experiments to measure the Mueller matrices of the real seawater channel at a distance of 10.1m, the results were the same as here, see Appendix for more details.

# 3. Experimental procedures and results of underwater QKD

We then carried out a proof-of-principle experiment for the polarization-based BB84 QKD protocol over a 2.37m water channel. To implement the polarization encoding BB84 protocol QKD, the transmitter (Alice) randomly encodes a single photon with one of four polarization states: horizontal, vertical, $45°$ linear polarization, or $135°$ linear polarization, denoted by $|H\rangle, |V\rangle, \frac{1}{\sqrt{2}}(|H\rangle+|V\rangle), \frac{1}{\sqrt{2}}(|H\rangle-|V\rangle)$, respectively.

It then transmits this to the receiver (Bob), who measures each arriving photon and attempts to identify the states Alice has transmitted. Alice and Bob share the sifted key after basis reconciliation and acquire the secure key after error correction and privacy amplification.

The QKD transmitter (Alice) in our experiment (Fig.

1) operated at a frequency of 1MHZ. Random numbers were generated by a pseudo random number generator on the Field-Programmable Gate Array (FPGA) control board. The random numbers, which were used to determine which of the four lasers would emit pulses, were sent to post-processing software in the computer. The laser diodes (with a wavelength of 450nm) were driven by Transistor-Transistor Logic (TTL) signals and the laser pulses were attenuated to 0.1 photon per pulse by adjusting the attenuators. Through the modulation of polarizers and half waveplates, whose rotation accuracy were controlled within 5 arc minutes, the polarization states of photons emitted from the four lasers were encoded. The four lasers are matched within 2mm at the transmitter. Non-polarizing beam splitters (NPBS) were introduced to decrease the influence of BS on polarization states. The quantum channel was a simulated seawater channel formed using a tank with a length of 2.37m. The simulated seawater was prepared in the same way as that in the Mueller matrix measurement experiment. To obtain the channels with different attenuation coefficients, we added different amounts of Al(OH)$_3$ to the simulated seawater. We prepared the Al(OH)$_3$ suspension by mixing 1.2g Al(OH)$_3$ and 170mL ultra-pure water. The suspension was introduced into the tank in increments of 15mL. Each time the suspension was added, the tank was thoroughly stirred, and then the attenuation and absorption coefficients were measured. At Bob's site, laser pulses passed onto a NPBS where they were randomly transmitted or reflected. Along the transmitted path, pulses' polarizations were analyzed by a polarizing beam splitter (PBS). Along the reflected path, pulses' polarizations were analyzed by a half waveplate and a polarizing beam splitter. Four single photon detectors (SPD) are used to detect the single photon pulses; the detection results were transmitted to a computer by the FPGA control board. The FOV of the detector was 8mrad and the bandwidth of the filter was 10nm, thus part of the background light can be filtered. The post-processing software in the computer performed basis reconciliation, error correction with the CASCADE algorithm[35], and privacy amplification.

The performance of QKD channel is characterized by quantum bit-error rate (QBER) and secure bit rate. In our work, we implemented the underwater BB84 protocol and measured these two important parameters for different channels. The QBER is defined as the ratio of wrong bits to the total number of bits received and the function reads[36]:

$$QBER = \frac{N_{wrong}}{N_{total}} = \frac{R_{wrong}}{R_{total}} \qquad (4)$$

where $N_{wrong}$ ($N_{total}$) is the wrong (total) number of bits received, and $R_{wrong}$ ($R_{total}$) is the wrong (total) number of bits received per second. In BB84 protocol, the sifted key rate is [36]:

$$k = f \cdot \mu \cdot T \cdot q \cdot \frac{\eta}{2}, \qquad (5)$$

where f is the pulse frequency of the laser, T is the transmission ratio of channel, q is the sifting factor which is usually $\geq 1$ and typically 1 or 1/2 , $\mu$ is the average photon number per pulse, and $\eta$ is the efficiency of the detector. In our experiment, the devices were installed in a dark room and $R_{wrong}$ mainly comes from the imperfect optical elements and dark counts. For QKD, the secure final key is extracted from the sifted key. When extracting the secure key from the sifted key, the extraction ratio is about 11% in our system.

Fig. 3a shows the absorption and attenuation coefficients of the simulated seawater as a function of the volume of Al(OH)$_3$ suspension; both coefficients increase nearly linearly[31]. To clearly demonstrate the performance of our underwater QKD channel, we firstly performed the QKD over the air channel for comparison. We obtained a QBER of 1.58% and a secure bit rate of 422.96 bits/s. This QBER mainly comes from background light, dark counts of the detectors, and imperfections in the optical components. Next, we measured the QBER and bit rate for different simulated seawater scenarios, as shown in Fig. 2a, the results are presented in Fig. 2b. Obviously, with an increase in the attenuation coefficient of the water channel from 0.11 to 0.68/m, the QBER increases from 1.65% to 3.5%, and the secure bit rate decreases from 337.2 to 37.9 bits/s. When the turbidity of seawater is low, the QBER and secure bit rate in seawater are closed to that in air channel. The main reason for the increase in QBER after adding Al(OH)₃ is that the decrease of total received photons leads to an increase in the proportion of background light and dark counts (i.e. an increase in QBER caused by backgroud light and dark counts). The other reason is the increase in QBER caused by the depolarization of seawater on photons. The depolarization is extremely small according to the slow upward trend of the QBER increase. And the bit rate can be further improved by enhancing the repetition frequency of the single photon pulses.

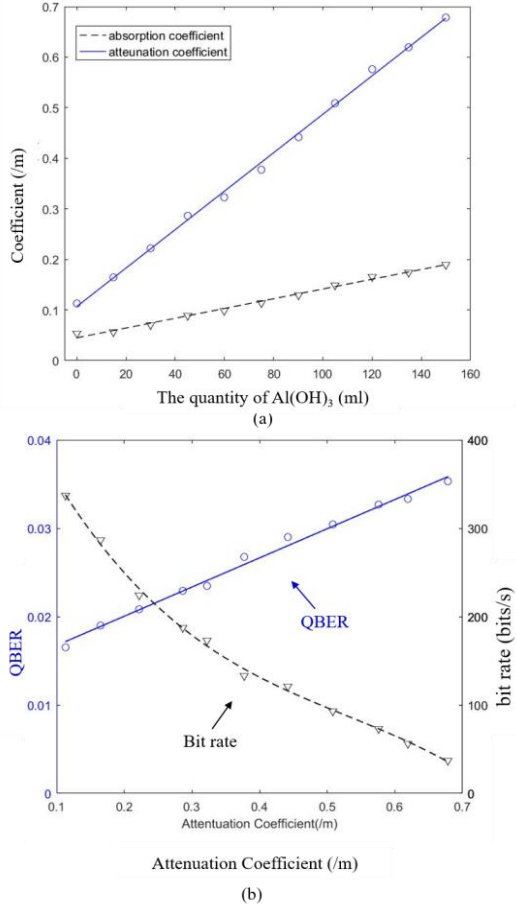

**Fig. 3.** (color online) (a): Attenuation coefficient (dashed line) and absorption coefficient (solid line) of the water channel as a function of Al(OH)₃ suspension (i.e., water turbidity). (b) Quantum bit error rate (QBER; left axis) and bit rate (right axis) along with the attenuation of the water channel. The blue circles and black inverted triangles represent experimental data.

## 4. Discussion

In our work, we have performed the first proof-of-principle experiment for the BB84 protocol QKD over the water channel. We have demonstrated that the Mueller matrix for the water channel is close to the unit matrix, which means that the seawater channel is polarization-preserving. The QBER is less than 3.5%, which is far smaller than the upper bound of the QBER (11%) for secure QKD [36, 37], for a 2.37m tank with different attenuation coefficients (0.11 to 0.68/m). As the attenuation would be lower for seawater farther away from the coastline, underwater QKD for longer transmission distances may be possible. For Jerlov Type I seawater with an attenuation coefficient of 0.03/m, the 3.5% QBER of underwater QKD can also be implemented at a distance of about 53.7m acquired according to the same attenuation with 0.68/m × 2.37m. According to the above results, we find that QKD over seawater can be performed with a low QBER.

As a proof-of-principle QKD over water channel, we implement the original BB84 protocol QKD. Some components and techniques have not been introduced, such as true quantum random number generator, decoy state, phase randomization between pulses, real-time error correction and privacy amplification. These components and techniques will ensure the security of underwater QKD. Next step, we will investigate underwater QKD with these components and techniques. Also, we will increase the pulse repetition rate and investigate underwater QKD with long distance. For outdoor QKD, reducing the background light is necessary to decrease the QBER, which can be implemented using a narrow bandwidth filter, small FOV and short gate time. Another potential approach to reducing background light for practical underwater QKD is to choose a wavelength corresponding to a Fraunhofer line. Besides, classical communication is needed in the post-processing of QKD. For practical polarization encoding QKD, the single laser fiber-based polarization encoding scheme with phase modulation can also ensure the security and the robustness of the system[38, 39]. In practical underwater QKD systems, Acquisition, Tracking, Pointing (ATP) technology is required to align the transmitter and receiver; as in laser communication between satellites and Earth stations, wandering of the transmitter and receiver, and beam dithering caused by turbulence, will inevitably occur. According to the unified theory of coherence and polarization of light beams, the completely polarized light will not be affected in Kolmogorov ocean turbulence[40, 41]. Even when the oceanic turbulence exists, the secure key can be generated in theory[42].

## 5. Conclusions

In summary, we analyzed the polarization preservation properties of the water channel by both experimental and simulated methods and experimentally demonstrated underwater QKD over a 2.37m simulated seawater channel. The polarization is nearly unchanged during light transmission over a water channel according to the Mueller matrix of the water channel. By using Monte Carlo simulation, we show that most of the received photons are unscattered and that the polarization of other photons changes little owing to the small scattering angles. We performed the polarization encoding BB84 protocol QKD over the water channel. The results show that low QBER can be obtained for different attenuation coefficients, further demonstrating the feasibility of underwater QKD with polarization encoding. For practical underwater QKD, the laser in the blue-green optical window of seawater is applicable because attenuation will be relatively low of electromagnetic spectrum underwater.

## 6. Appendix

To verify the feasibility of practical underwater QKD, we also investigated the polarization preservation properties of the real seawater channel at a distances of 10.1m. The seawater was collected from Shazikou, Qingdao (N36◦06041.1500, E120◦32042.9900); the attenuation coefficient, absorption coefficient, and salinity were 0.997/m, 0.273/m, and 3.55%, respectively. We changed the intensity of the laser pulses to 1 mw, and detected the pulses with a polarization meter (PAX1000), shown in Fig. 4, which can acquire the stokes vectors S0 (over air channel) and S1(over water channel). In order to construct the system of 16 linear equations, we generated four kinds of polarization states S0: horizontal, vertical, 45◦ linear polarization, and right circular polarization, by rotating the polarizer and quarter waveplate in the transmitter (the quarter waveplate was removed for the three linear polarization states). In the experimental system, the background light and most of the scattered photons were effectively filtered by a 3-mm effective diameter and a 2◦ field of view (FOV) in the receiver; the ratio of the intensity of background light to that of signal light was less than 1/100. The detected Mueller matrix is shown in Fig. 5, it is close to the unit matrix. The nearly unit Mueller matrix reflects the good polarization preservation properties for the water channel on light.

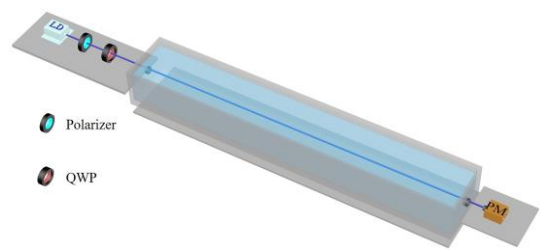

**Fig. 4**. (color online) Sketch of polarization maintaining the experimental system of the water channel. QWP is a quarter waveplate and PM is a polarization meter.

The minimum value of the diagonal elements is larger than 0.964, the maximum absolute value of the non-diagonal elements is smaller than 0.0288 for the measured Mueller matrices, and the fidelity is 0.973±0.0151. For the polarization encoding BB84 protocol, the QBER caused by the channel was calculated according to Eq.4. Here, the QBER caused by dark counts and imperfect photon sources were not studied because $N_{total} \approx N_{signal}$[22, 36], where Nsignal is the number detected by Bob. The Nwrong was obtained according to the stokes vector, whose first element represents the intensity, of photons reaching the detector. The QBER caused by the depolarization of seawater channel according to the measured Mueller matrices is about 2.7%, far smaller than the up boundary of the QBER for secure QKD. Thus, underwater QKD in real seawater channel is feasible. We also measured the Mueller matrix of the water in Section 2 using PAX1000, the results show that the results of these two methods are almost the same.

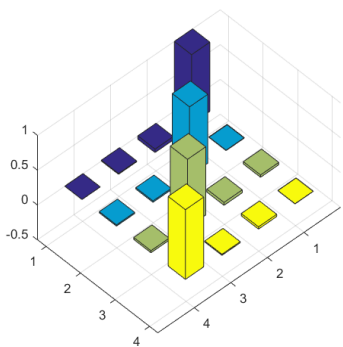

**Fig. 5.** (color online) Mueller matrix of the real seawater channel with a distance of 10.1m

## Acknowledgments

This work was supported by the National Natural Science Foundation of China (Grant No. 61575180, 61701464 and 11475160). ShiCheng Zhao and WenDong Li contributed equally to this work